# Synthesis and photoemission study of as-grown superconducting $MgB_2$ thin films

K Ueda, H. Yamamoto, M. Naito

NTT Basic Research Laboratories, NTT Corporation,

3-1 Wakamiya, Morinosato, Atsugi, Kanagawa 243-0198, Japan

**Abstract**

As-grown superconducting thin films of $MgB_2$ were prepared by molecular beam epitaxy (MBE), and studied by X-ray and ultraviolet photoelectron spectroscopy (XPS and UPS). Only films prepared at temperatures between 150 and 320 showed superconductivity. The best $T_C^{onset}$ of 36 K was obtained with a sharp transition width of 1 K although the film crystallinity was poor. The in-situ photoelectron spectra obtained on the surfaces of the MBE grown $MgB_2$ films were free from dirt peaks. The XP spectra revealed the binding energy of the Mg 2p levels in $MgB_2$ is close to that of metallic Mg and the binding energy of B 1s is close to that of transition-metal diborides. The valence UP spectra showed a clear Fermi edge although the density of states at $E_F$ is low and the major components of the valence band are located between 5 and 11 eV.

*Corresponding author.

Kenji Ueda

Postal address: NTT Basic Research Laboratories, 3-1 Wakamiya, Morinosato,

Atsugi, Kanagawa 243-0198, Japan

Phone: +81-46-240-3356

Fax: +81-46-240-4717

E-mail address: kueda@will.brl.ntt.co.jp

## 1. Introduction

The recent discovery of superconductivity at 39 K in $MgB_2$ [1] has generated great scientific interest. $MgB_2$ has the highest superconducting transition temperature ($T_C$) among non-oxide materials (except for some doped fullerenes), and the $T_C$ is close to that of $La_{2-x}Ba_x$(or $Sr_x$)$CuO_4$, which were the first high-$T_C$ superconductors. Hence, it is important to understand the origin of the superconductivity of $MgB_2$, especially in relation to cuprate superconductors. The $T_C$ is slightly higher than the theoretical upper limit predicted for phonon-mediated superconductivity [2], which had been widely accepted until the discovery of the cuprate superconductors. However, [11]B-NMR studies [3] and the B isotope effect [4] showed that $MgB_2$ is a conventional BCS superconductor.

Accurate information on its electronic structure is a prerequisite for understanding of physics in $MgB_2$. There have been several photoemission studies that investigated the electronic structure of $MgB_2$ [5, 6]. However, these studies were all undertaken using sintered samples and/or annealed films. The sintered specimens included some oxidized Mg on their surface as suggested by Vasquez et al [5]. Since photoemission spectroscopy is a highly surface-sensitive method, the preparation of a clean surface is very important as regards clarifying the intrinsic nature of the superconducting phase. In this work, we prepared as-grown superconducting $MgB_2$ thin films by molecular beam epitaxy (MBE) and investigated the films in-situ using X-ray and ultraviolet photoelectron spectroscopy (XPS and UPS).

## 2. Experimental

The $MgB_2$ films were grown on various substrates, namely $SrTiO_3$ (001), sapphire R, sapphire C, and H-terminated Si (111) in a custom-designed MBE chamber (basal pressure of 1 2 $10^{-9}$ Torr) from pure metal sources using multiple electron beam evaporators [7, 8]. The evaporation beam flux of each element was controlled by electron impact emission spectrometry (EIES) via feedback loops to the electron guns. We described the details of the synthesis of the superconducting $MgB_2$ thin films in our previous paper [9]. The crystal structure was characterized by X-ray diffraction (XRD: 2 - scan) and reflection high-energy electron diffraction (RHEED). We measured the resistivity using the standard four-probe technique. The compositions of some films were examined by inductively coupled plasma spectrometry (ICP) analysis. For XPS and UPS, the grown films were transferred in vacuo to an analysis chamber with a base pressure of 3 $10^{-10}$ Torr. The spectrometer was a VG-CLAM2 and the light sources were VG XR3E2 (Mg K ) and VG UVL-HI (He I, He II). We obtained the spectra at an ambient temperature with the photoemission normal to the surface of the film.

## 3. Results and Discussion

The main problem regarding the growth of $MgB_2$ films is the high volatility of Mg. The films we prepared above 350 were significantly deficient in Mg, and semiconducting with no trace of superconductivity. Superconductivity was observed only for films prepared at growth temperatures between 150 and 320 [9]. The films we prepared below 100 were again semiconducting. Figure 1 shows a typical temperature dependence of resistivity of as-grown superconducting $MgB_2$ films. The $T_C$ is appeared at around 32 ñ 33 K. This specific film was grown on sapphire C at $T_S$=300. The temperature dependence of the resistivity is much weaker (smaller residual resistivity ratio) than that for single crystals. Our best $T_C$ was 36 K with a transition width of 1 K.

As regards film crystallinity, we observed no XRD peak for the films on the $SrTiO_3$ (001) and sapphire R substrates (square or rectangular surface). These films had halos, and sometimes rings, in RHEED. By contrast, we were able to observe faint peaks for the films corresponding to $MgB_2$ (00$l$) for H-Si (111) and sapphire C substrates (hexagonal surface). The films also showed spots in RHEED, indicating a single crystalline nature. These results show that films grown on $SrTiO_3$ (001) and sapphire R are amorphous or polycrystalline, while films on H-Si (111) and sapphire C have a c-axis preferred orientation although their crystallinity is poor. For this reason, we chose films on sapphire C for our photoemission measurements.

Figure 2 shows the Mg 2p XP spectrum of the $MgB_2$ film on sapphire C. The arrows indicate the reported peak positions of Mg 2p of in $MgAl_2O_4$, $MgF_2$ and Mg metal [5, 10]. The peak position in our $MgB_2$ film is 49.5 eV. This is close to that of Mg metal (49.8eV). However, it could not lead to any definitive conclusion on Mg valence because the chemical shift shows no systematical trend in Mg XPS [10]. It should be noted that the in-situ Mg 2p spectrum of the as-grown $MgB_2$ film has a single component except a shoulder. This is in contrast with the experiment reported by Vasquez et al. on a chemically etched post-annealed film [5], where they also observed another component originating from $MgCO_3$ and/or $Mg(OH)_2$ as a peak around 51-52 eV.

Figure 3 shows the B 1s XP spectrum of the same film. Again the arrows denote the reported binding energy of B 1s in $B_2O_3$, BN, B and transition metal diborides (boride), respectively [5, 10]. The B 1s peak is observed at 187.1 eV, which is within the binding energy variation range for typical transition metal diborides (187.1-188.3eV [5, 10-11]). This observation is qualitatively consistent with reported calculations, and MEM (Maximum Entropy Method) demonstrating that B-B bonding is two-dimensionally covalent ($sp^2$ covalent) [12-15].

The valence UP spectra of the $MgB_2$ film are shown in figure 4 (a). The ultraviolet source is He

I (21.2eV) for lower spectra and He II (40.8 eV) for upper spectra. Our UP spectra showed a clear Fermi edge although the total density of states (DOS) at $E_F$ is low. The major components of the valence band are located between 5 and 11 eV. These features agree qualitatively with the experimental results reported by Vasquez et al. [5] although their spectra were obtained with a much higher photon excitation energy (1486.6 eV, Al K ) and on chemically etched post-annealed films. Neither our experimental spectra nor those obtained by Vasquez et al. support the theoretical band calculation that predicts a pile-up of DOS with a dominantly B 2p character at ~ 2.5 eV [12]. The low DOS at $E_F$ seems to be consistent with specific heat experiments indicating that the linear coefficient ( ) of the electronic specific heat is rather small ( 3 mJ/mol K$^2$) [4]. This implies that the high-$T_C$ superconductivity in $MgB_2$ is not due to a high DOS at $E_F$. Finally, it should be noted that $MgB_2$ spectra have some similarity with those obtained for high-$T_C$ cuprates (Fig. 4 (b) [7, 16]), in that the DOS at $E_F$ is low and the major component is well below $E_F$.

## 4. Summary

We prepared as-grown superconducting thin films of $MgB_2$ by MBE and studied them using photoemission spectroscopy. The films formed at temperatures between 150 and 320 showed superconductivity despite their poor crystallinity. In-situ XPS revealed that the binding energy of B 1s level in $MgB_2$ is close to that in transition-metal diborides, indicating that B-B bonding is sp$^2$ covalent. The binding energy of the Mg 2p is close to that in metallic Mg, but the valence of Mg cannot be determined conclusively. The valence spectra showed a clear Fermi edge although the total DOS at $E_F$ is low. The major components of the valence band are located between 5 and 11 eV, which agrees well with the XPS results reported by Vasquez et al.

**Acknowledgements**

The authors thank Dr. S. Karimoto and Dr. H. Sato for fruitful discussions and Dr. H. Takayanagi and Dr. S. Ishihara for their support and encouragement throughout the course of this study.

**Figure captions**

Fig. 1: The resistivity versus temperature curve of $MgB_2$ film on sapphire C at a substrate temperature of 300 . The inset is an enlargement around $T_C$.

Fig. 2: Mg 2p spectrum of the as-grown $MgB_2$ superconducting thin film. Allows indicated the positions of the binding energy of the Mg 2p level in $MgAl_2O_4$, $MgF_2$ and Mg metal.

Fig. 3: The B 1s spectrum of as-grown $MgB_2$ superconducting film. Each arrow indicates the position of the binding energy of B 1s in $B_2O_3$, BN, B and transition metal diborides (boride).

Fig. 4: (a) Valence UP spectra of $MgB_2$ film. The ultraviolet source is He I (21.2eV) for lower spectra and He II (40.8 eV) for upper spectra. The inset is an enlargement of the UP spectra around $E_F$. (b) Valence UP spectra of MBE-grown $La_{1.85}Sr_{0.15}CuO_4$ (LSCO) and $La_{1.89}Ce_{0.11}CuO_4$ (NCCO) films. The ultraviolet source is He I.

K. Ueda, H. Yamamoto, M. Naito

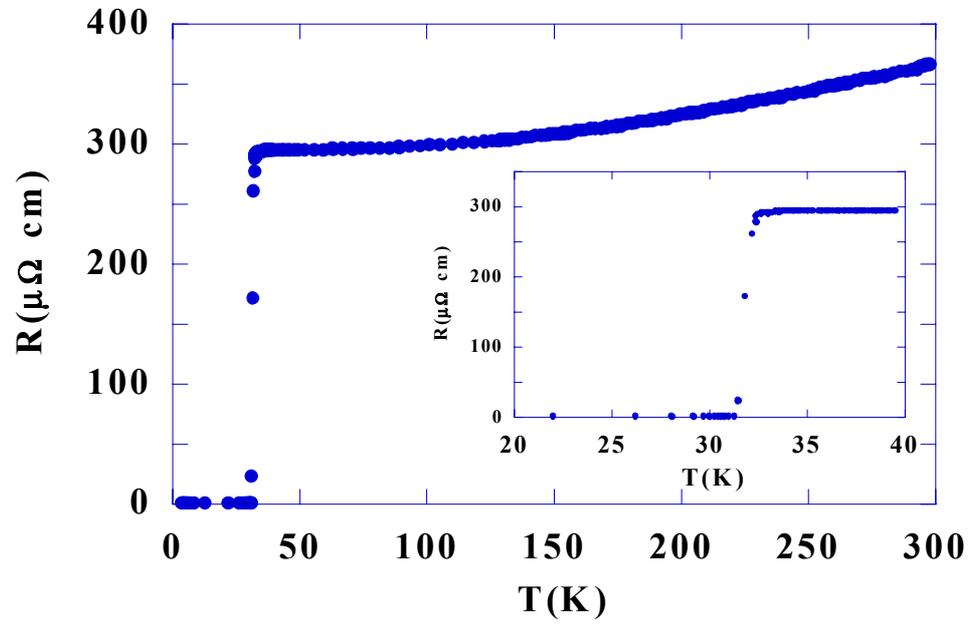

**Fig. 1/PCP-3/ISS2001**

K. Ueda, H. Yamamoto, M. Naito

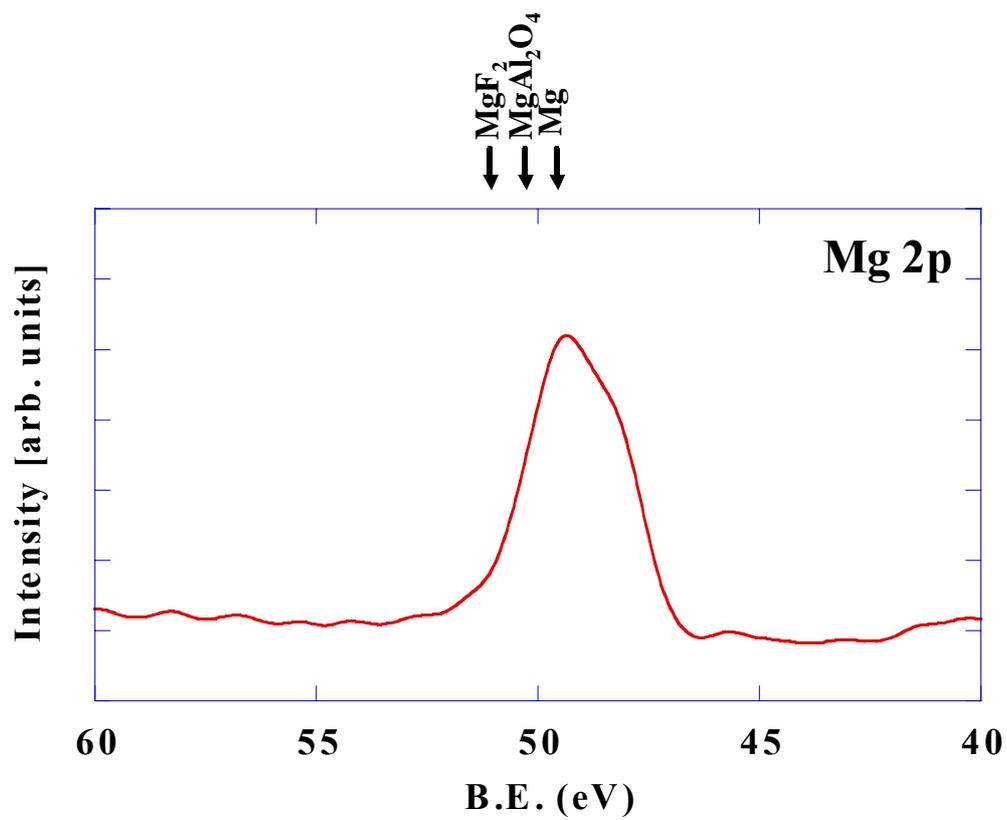

K. Ueda, H. Yamamoto, M. Naito

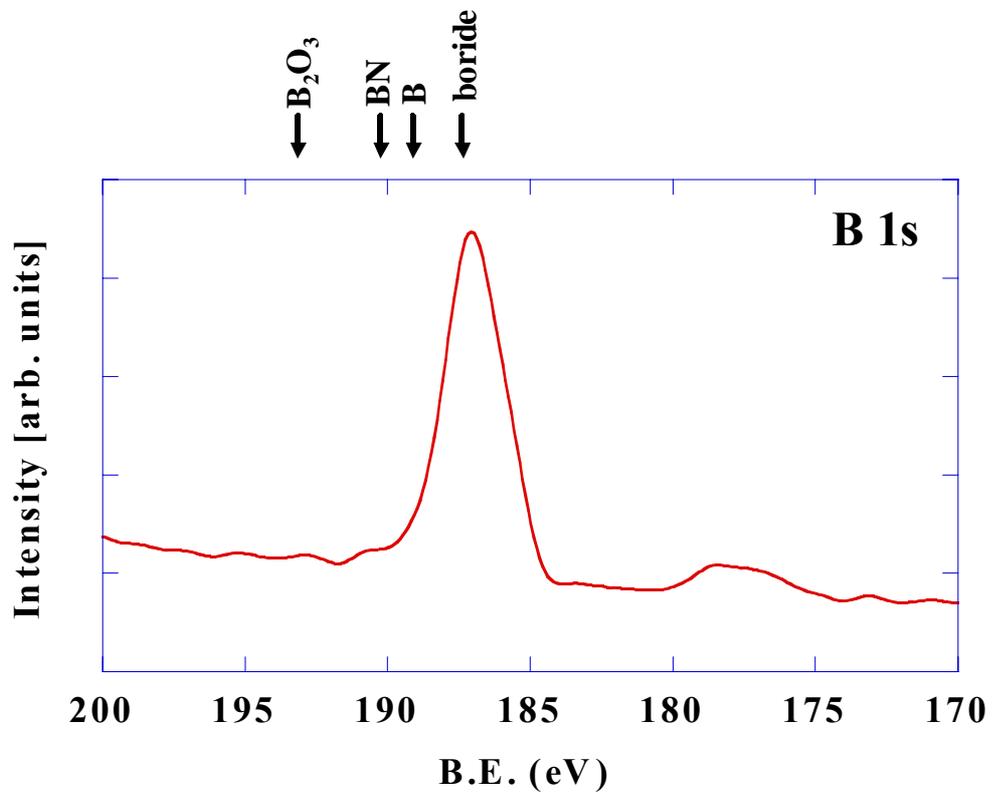

Fig. 3/PCP-3 /ISS2001

K. Ueda, H. Yamamoto, M. Naito

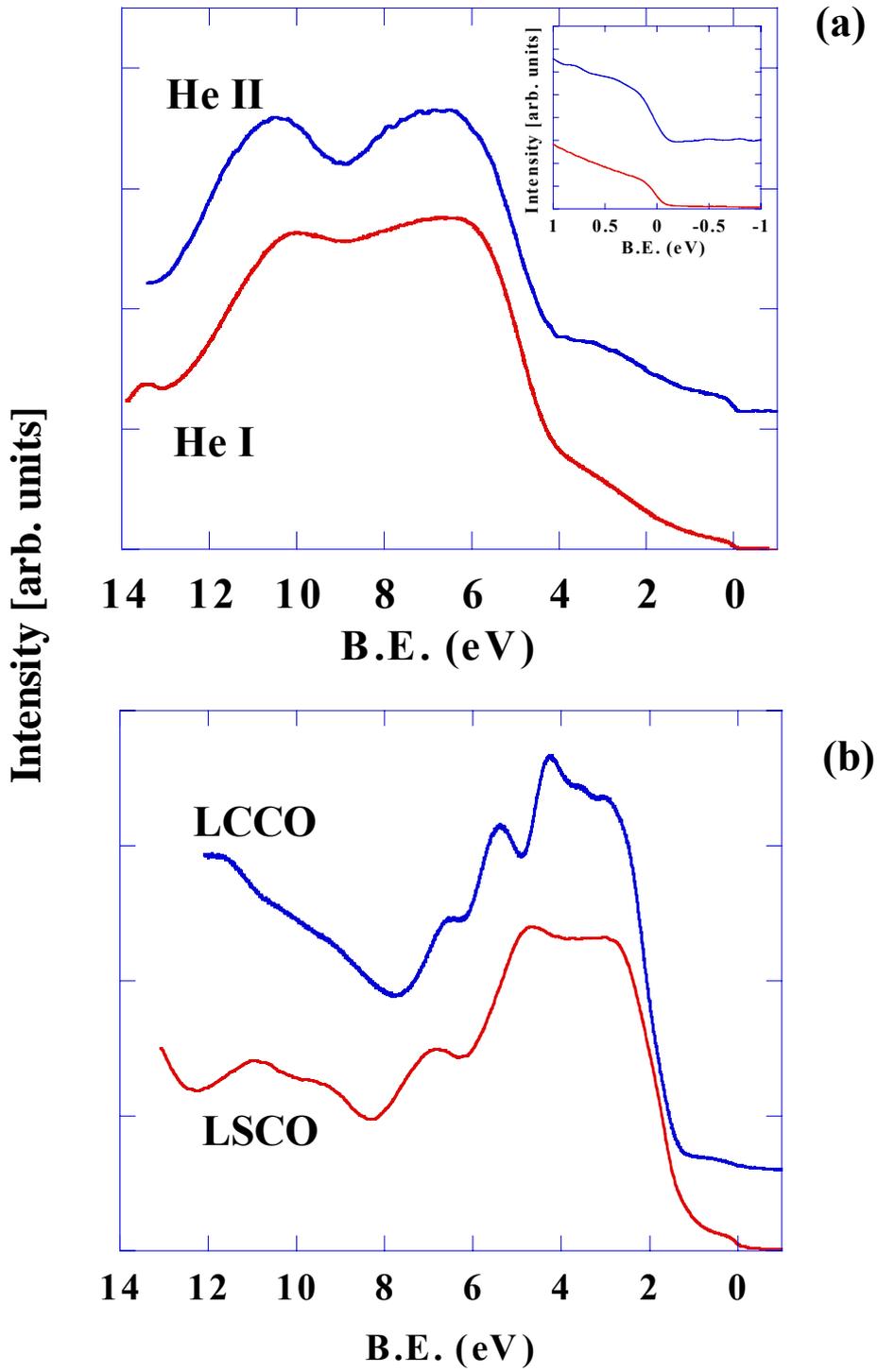

Fig. 4/PCP-3 /ISS2001